\documentclass[12pt,american]{article}


\usepackage{graphicx}
\usepackage{amsmath, amssymb, amsthm}
\usepackage{natbib}

\usepackage[usenames,dvipsnames,svgnames,table]{xcolor}

\usepackage[mathscr]{euscript} 
\usepackage{bbm}

\usepackage{enumitem}
\setlist[enumerate]{itemsep=5pt,topsep=3pt}
\setlist[itemize]{itemsep=2pt,topsep=3pt}
\setlist[enumerate,1]{label=\arabic*.}

\renewcommand{\phi}{\varphi}
\renewcommand{\epsilon}{\varepsilon}

\usepackage{mdwlist}

\usepackage[citecolor=blue, colorlinks=true, linkcolor=blue]{hyperref}

\usepackage{caption}
\usepackage{subcaption}

\usepackage{algorithmic}
\usepackage{algorithm}

\usepackage[left=1.in, right=1.in, top=1.in, bottom=0.9in, includehead, includefoot]{geometry}

\theoremstyle{plain}

\theoremstyle{definition}

\usepackage[multiple]{footmisc}

\usepackage{authblk}

\title{Does Foreign Debt Contribute to Economic Growth?
	\thanks{We would like to thank seminar participants at Keio University, Korea University, Kwansei Gakuin University, Peking University, Sichuan University, and Waseda University and acknowledge valuable comments from Laura Alfaro, Ippei Fujiwara, Wentao Fu, Munechika Katayama, Jinill Kim, Takuma Kunieda, Jong-Wha Lee, Cheolbeom Park, Kwanho Shin, Kozo Ueda, Yong Wang, Lei Zhang and Lijung Zhu, and financial support from Korea University (K2009811 and K1922081) and BrainKorea21 Plus (K1327408 and T192201). Corresponding author: Tomoo Kikuchi. Nishi-Waseda Bldg.7F, 1-21-1 Nishi-Waseda, Shinjyuku-ku,
		Tokyo 169-0051 Japan. Email: \texttt{tomookikuchi@waseda.jp}.}
}

\author[a]{Tomoo Kikuchi}
\author[b]{Satoshi Tobe}

\affil[a]{\small School of Asia-Pacific Studies, Waseda University}
\affil[b]{\small School of Policy Studies, Kwansei Gakuin University}

\begin{document}
	
\maketitle
	
\begin{abstract} 
	
\noindent	We study the relationship between foreign debt and GDP growth using a panel dataset of 50 countries from 1997 to 2015. 
We find that economic growth correlates positively with foreign debt and that the relationship is causal in nature by using the sovereign credit default swap spread as an instrumental variable. 
Furthermore, we find that foreign debt increases investment and then GDP growth in subsequent years. Our findings suggest that lower sovereign default risks lead to higher foreign debt contributing to GDP growth more in OECD than non-OECD countries.
	
\vspace{1ex}


\noindent \textbf{JEL\ Classification:} F21; F34; O16

\end{abstract}

\clearpage

\section{Introduction}

\begin{figure}[ht!]
	\begin{center}
		\centering
		\includegraphics[width=.6\textwidth]{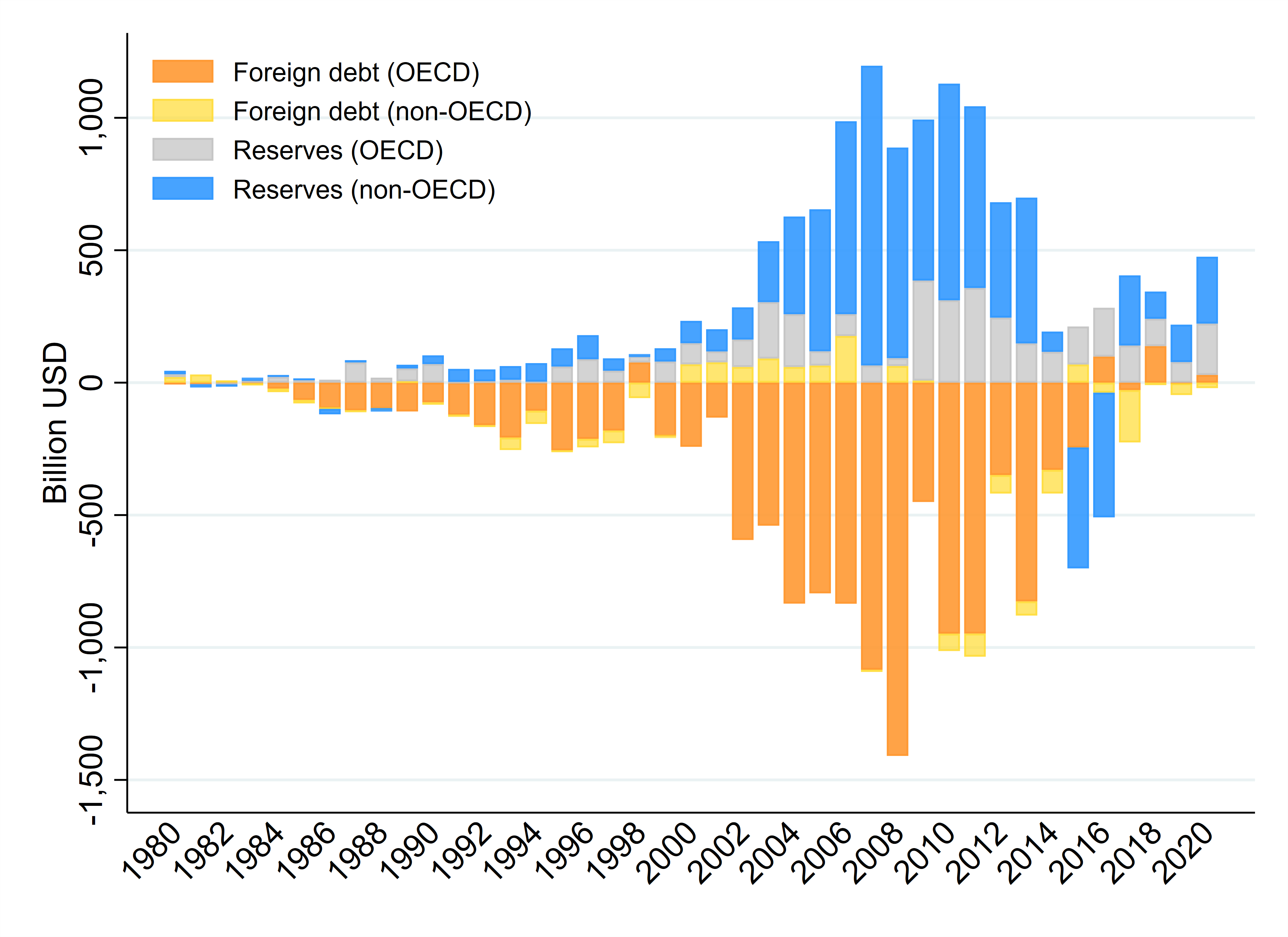}
		\caption{The Sum of Net Foreign Debt and Foreign Reserves for about 200 Countries from 1980 to 2020.}
		\label{fig:Foreign_debt_vs_reserves_rev}
	\end{center}	
	\footnotesize Note: Foreign debt represents the debt instrument of portfolio investment. This figure plots aggregated series of all reporting countries in the balance of payment statistics provided by IMF. Data of China's foreign debt are missing for 2016-2020.
\end{figure}

OECD countries have issued most of debt in the world despite growing slower than the rest of the world.
In fact, 98 percent of  
net foreign debt in the world between 1980 and 2020 (12.5 trillion USD) was issued by OECD countries (see Figure \ref{fig:Foreign_debt_vs_reserves_rev}) although their GDP grew only 2.5 times while GDP of the rest of the world grew 8.0 times in the same period.
Moreover, most of debt issued by OECD countries are financed by other OECD countries.\footnote{Bilateral portfolio investment data show that 91 percent of debt investment in OECD countries originated from other OECD countries
between 1997 and 2020. Source: CPIS, IMF}
This might not be surprising given the size of OECD economies even though their share in the world GDP has fallen from 84 percent to 62 percent in the period.

If capital is not flowing more to countries that are growing faster, 
there must be some financial frictions that do not allow capital to flow freely to where the return is the highest or there must be reasons other than returns on capital that matter for investors. 
Indeed, when international investors invest in foreign debt, they care about risks as much as about returns. One of the most important risks for investing in foreign debt is the sovereign default risk. 
Therefore, other things being equal, the higher the sovereign default risk of a country is, the lower must be the international investment in foreign debt in that country. The question remains how foreign debt then contributes to economic growth.

Market perception about the likelihood of sovereign default is reflected in the sovereign credit default swap (CDS) spread, which is a derivative  investors buy to hedge against the sovereign risk. By using the CDS spread as an instrumental variable (IV) 
this paper investigates the causal relation between net foreign debt (henceforth simply ``foreign debt'') and GDP growth using a panel dataset covering 50 countries from 1997 to 2015. 
We argue that the predicted values in the first stage of panel IV regression can be interpreted as changes in foreign debt caused by exogenous supply-side shocks. 
We find that foreign debt increases GDP growth when we exclude major offshore financial centers and oil exporters. We justify this exclusion on the ground that those countries, although recording a very high volume of international financial transactions, might distort our analysis because they channel a very small portion of capital inflows into domestic investment - the channel we explore in this paper. 
In fact, we find that foreign debt increases investment and GDP growth and the dynamic effect lasts for a few years by employing the local projection analysis developed by \cite{jorda2005estimation}. This indicates that foreign debt contributes to GDP growth through capital accumulation.
Based on the coefficient we obtain from our model we calculate the contribution of foreign debt to economic growth in OECD and non-OECD countries. 
We find that foreign debt increases GDP growth in OECD countries (0.155 percentage point)  60 percent more than in non-OECD  countries (0.096 percentage point). 
Hence, the novel channel between foreign debt and growth we identify via the CDS spread points to the ``real'' effects of the ``exorbitant privilege'' the US and some other OECD countries have \citep[c.f.,][]{gourinchas2007world}.
We conclude that lower sovereign default risks lead to higher foreign debt contributing to GDP growth in OECD countries more than in non-OECD countries.

In the neoclassical model the law of diminishing returns implies that capital - ceteris paribus - would flow from rich countries where the marginal product of capital is low to poor countries where it is high.
In other words, capital should flow to countries that grow faster. 
The observation that capital does not flow from rich to poor countries as the neoclassical model predicts is known as Lucas' paradox based on \cite{lucas1990doesn}. Since Lucas' seminal work, the literature has decomposed capital flows in the financial account to examine the direction of each flow and its effects on economic growth. It is well-established that foreign direct investment (FDI) contributes to economic growth in developing countries with a high level of human capital in \cite{borensztein1998does} or developed financial markets in \cite{alfaro2004fdi}. Their finding seems to be consistent with the neoclassical model. On the other hand, there is evidence that ``upstream capital flows'' - capital flows from poor to rich countries - are caused by fast-growing emerging economies that accumulate foreign reserves in the US and other rich countries.  \citep[e.g.,][]{gourinchas2007world,mendoza2009financial,gourinchas2013capital,alfaro2014sovereigns}.\footnote{In theoretical models with no FDI, financial frictions may limit domestic investment and lead to upstream (portfolio investment) flows from ``poor'' countries to ``rich countries''  and sustain the cross-country inequality when financial markets are integrated globally \citep[e.g.][]{matsuyama2004financial,kikuchi2009endogenous,kikuchi2018volatile}. If FDI is allowed, investors in rich countries could circumvent host-country financial frictions and invest in poor countries.}

Our paper is closely related to the papers above but has a different focus. 
Instead of focusing on the causes of upstream capital flows we investigate how foreign debt, which is predominantly issued by OECD countries, contributes to GDP growth without distinguishing between ``upstream'' and ``downstream'' flows.
In fact, there is a priori no reason why there should be differential effects of capital flows on GDP growth depending on the source.  
Moreover, foreign reserve accumulation by non-OECD countries that constitutes upstream capital flows is only 
a small part of investment in foreign debt in OECD countries. 
Since net foreign debt of OECD countries  largely mirrors foreign reserve holdings of non-OECD countries as shown in Figure \ref{fig:Foreign_debt_vs_reserves_rev},\footnote{The sum of net foreign debt across all countries in the world must be zero except for the discrepancies that foreign reserves create; there is no net concept for foreign reserves and every foreign reserve holding must have its counterpart recorded in foreign debt as its transactions typically involve buying or selling debt securities. For example, China's foreign reserves held as the US treasury securities are counted as foreign debt for the US.}
the effect of foreign debt on GDP growth we investigate includes the effects of foreign reserve holdings on GDP growth in the reserve-issuing country.

The positive relationship between foreign debt and GDP growth we find is the opposite of what others find for developing countries. For example, \cite{pattillo2002external} finds that the effect of foreign debt on per capita growth is negative among average developing countries. Similarly,  \cite{clements2003external} finds that
high levels of debt can depress economic growth in low-income countries.
Those findings are in line with debt overhang problems in low-income economies where a country's debt service burden to foreign lenders is so heavy that it creates disincentives to invest in the country \citep[see][]{krugman1988financing}. We differ from this line of literature in that 1) our sample includes low-income as well as high-income countries; 2) our IV approach controls for the incentives to invest in a country; and 3) our battery of controls such as regulatory quality and the government debt to GDP ratio accounts for the credibility of debtors. 
In other words, we control for unfavorable conditions in countries that suffer from debt overhang problems to analyze the effect of foreign debt on growth. 

Our work may also be contrasted with the literature that finds a negative relationship between public debt and growth \citep[see][]{reinhart2010growth}. Examining different samples of countries and periods, most works in the literature confirm a negative relationship between debt and growth \citep[see][for a survey]{reinhart2012public}. Our work differs from them as our main independent variable is  foreign debt, i.e., debt that a country owes to foreigners (both public and private entities).  Given home bias as in \cite{feldstein1979domestic}  foreigners are more responsive to shocks in international capital markets than domestic residents are. In other words, international investors are more selective than domestic ones, and therefore issuing debt domestically is fundamentally different from issuing debt internationally. Those differences, which we capture implicitly through our controls and IV approach, must be responsible for the positive relationship between foreign debt and growth.

The rest of the paper is organized as follows.
Section \ref{sec:growth-debt} describes the panel data and reports the main results. 
Section \ref{sec:mechanism} shows the contribution of foreign debt to economic growth in OECD and non-OECD countries.
Section \ref{sec:investment} investigates the investment channel and
Section \ref{sec:dynamic-effects} the dynamic effects of foreign debt. 
Section \ref{sec:conclusion} concludes.

\section{GDP Growth and Foreign Debt\label{sec:growth-debt}}
\subsection{Panel Data}
This section describes our panel data.
The data include several key variables: GDP growth, investment growth, foreign debt and the sovereign credit default swap (CDS) spread.
The following model uses real GDP growth as the key dependent variable.
GDP is measured in a constant local currency unit and provided in the World Development Indicator (WDI) database by the World Bank.
We also use investment growth as a dependent variable in our additional analysis in Section \ref{sec:investment}.Total investment is measured in a constant local currency unit and provided in the WDI database. Total investment can be decomposed into private and public investment, each measured in a constant US dollar unit rather than a local currency unit
due to data availability. The series are provided by the IMF.\footnote{We confirmed that main results hold when we use real GDP or investment growth measured in a constant US dollar unit.
However, we mainly use the series measured in a local currency unit to eliminate the valuation effects caused by changes in exchange rates as much as possible.}

The key right-hand-side variable is foreign debt normalized by GDP. Throughout this paper foreign debt denotes the debt instrument of portfolio investment, which captures international transactions of corporate and government debt provided in the International Financial Statistics by the IMF.
The series measure net capital inflows (gross inflows minus gross outflow). The balance of payments statistics also reports the debt instrument in a sub-category of other investment flows. We do not include the series in foreign debt as it mainly captures cross-border banking activities such as bank lending and deposit transactions,
which is not directly related to investment.

	\begin{table}[ht!]
	\centering
	\includegraphics[width=1\textwidth]{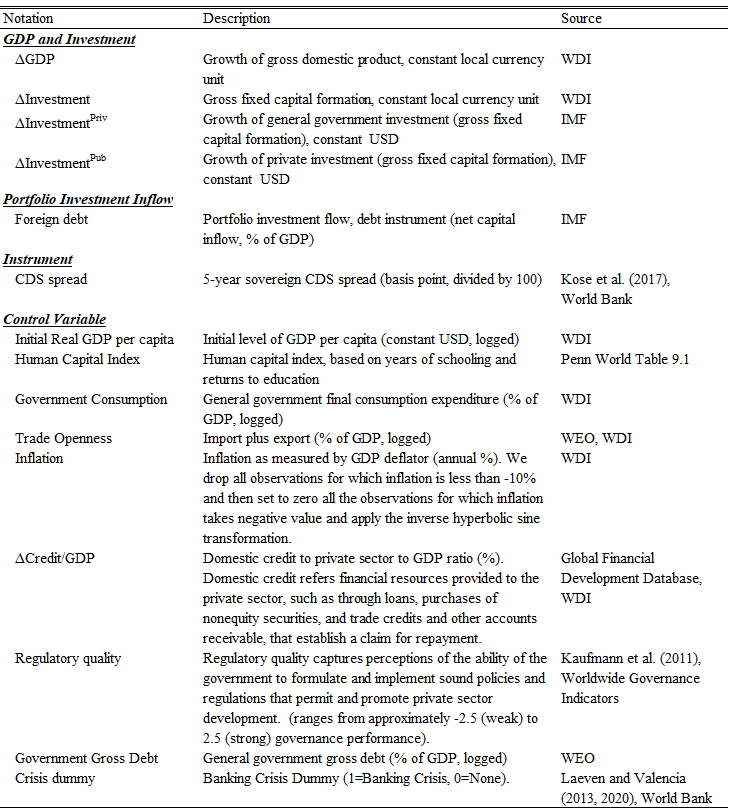}
	\caption{Definition and Notation of Variables.}
	\label{fig:Notation}
	\end{table}

\begin{figure}[ht!]
		\begin{center}
			\centering
			\includegraphics[width=.6\textwidth]{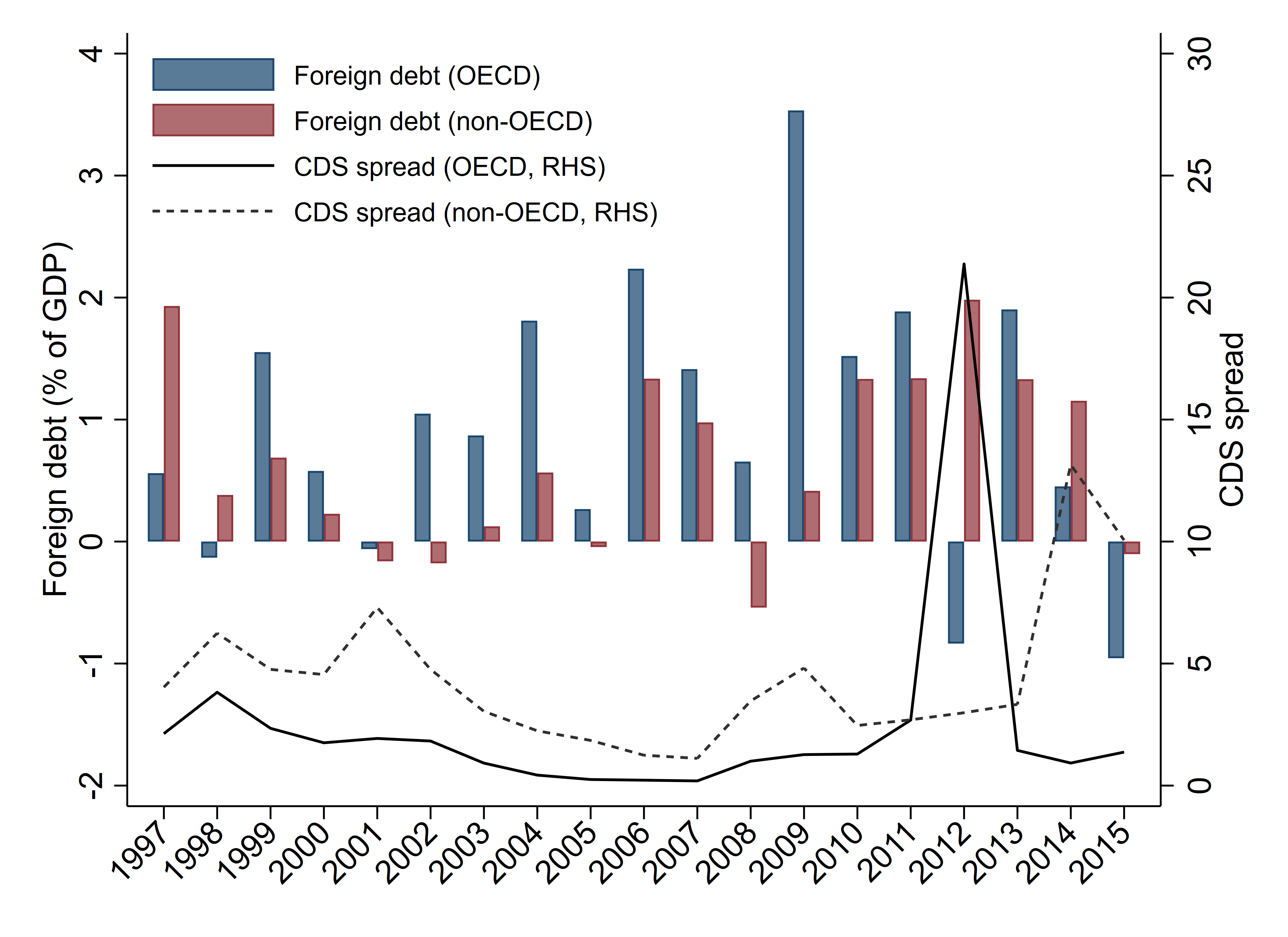}
			\caption{Foreign Debt and the sovereign CDS spread in OECD and non-OECD countries.}
			\label{fig:foreign_debt_cds}
		\end{center}	
		\footnotesize Note: This figure plots the average series of foreign debt and the CDS spread in OECD and non-OECD countries.
	\end{figure}

In the next section, we use the CDS spread as an instrument for foreign debt
to reduce the concern on the endogeneity between foreign debt and GDP growth.
The CDS  spread is the cost for investors to hedge against the sovereign default risk of a country.
Thus, it correlates with foreign debt.
For example, an increase in the CDS spread would lead to a decrease in foreign debt in a country because investors would reduce demand for securities that are considered more risky.
The data for the CDS spread are available from 1997 to 2015 provided by \cite{kose2017cross} and the World Bank. Figure \ref{fig:foreign_debt_cds} shows time series of net foreign debt as a percentage of GDP and the CDS spread in OECD and non-OECD countries from 1997 to 2015.
We can see that net foreign debt as a percentage of GDP is most of the time larger in OECD than non-OECD countries and that the CDS spread is lower in OECD than non-OECD countries except for 2012 when the debt crisis hit Europe.
Moreover, the CDS spread seems to be negatively correlated with net foreign debt in both country groups. 

We include controls standard in the growth regression literature such as the initial level of real GDP per capita, the human capital index, government consumption, trade openness, and the inflation rate \citep[e.g.][]{arcand2015too,durlauf2005growth}.
The details are summarized in Table \ref{fig:Notation}.
	
\begin{table}[ht!]
	\begin{center}
		\centering
		\includegraphics[width=.95\textwidth]{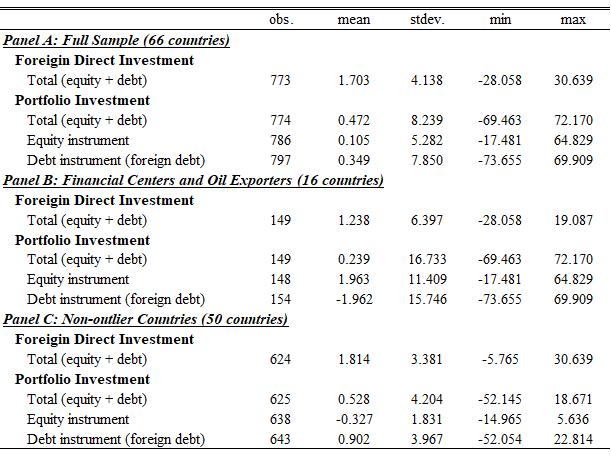}
		\caption{Summary Statistics of Outlier and Other Countries.}
		\label{fig:Summary_statistics2}
	\end{center}	
	\footnotesize Note: This table reports the summary statistics of 66 countries covering from 1980 to 2015. Table \ref{fig:country_list} summarizes the list of countries.
\end{table}

Our panel data are unbalanced covering 66 countries from 1997 to 2015 	because of the limited availability mainly of the CDS spread data. We also exclude two groups of countries from our sample. The first group contains offshore financial centers that record large transactions in international financial markets but typically re-invest capital inflows in third countries. 
The second group contains oil exporters that invest their oil revenues in foreign equity or debt and record large current account surplus (i.e., net capital outflow).
For both groups, capital inflows are not a source for domestic investment, which is the channel we would like to explore. We identify 9 countries as offshore financial centers following \cite{lane2018external} and 7 countries as major oil exporters, whose time-series average of the oil rents to GDP ratio exceeds 8 percent. The identified countries are listed in Tables \ref{fig:country_list} and \ref{fig:Ranking_oil} in Appendix \ref{sec:appendix}.

Table \ref{fig:Summary_statistics2} reports the summary statistics of FDI and portfolio investment showing that the offshore financial centers and oil exporters are indeed outliers. 
In the full sample reported in panel A, the standard deviation, minimum and maximum of portfolio investment are much larger than those of FDI. 
For example, the standard deviation of portfolio investment is 8.239, which is two times larger
than that of FDI (=4.138).
Panels B and C show that the large values of portfolio investment are attributable to the offshore financial centers and oil exporters, who generate noisy observations and cause inaccurate results. 
Moreover, panel C shows that total portfolio investment and foreign debt have a mean, a standard deviation, a minimum and a maximum that are similar but different from those of equity instrument.
This suggests that foreign debt is the key driver of portfolio investment covering 50 countries (excluding the outlier countries) from 1997 to 2015.
The 50 countries represent 48 percent of world GDP on average.
Table \ref{fig:country_list} reports the list of countries.

\subsection{The Panel Instrumental Variable Approach\label{sec:IV}}	

This section presents our main result.
We employ an instrumental variable (IV) approach to reduce concerns on endogeneity such as the reverse causality between foreign debt and GDP growth.
For example, higher GDP growth may lead to higher foreign debt because of higher demand for capital inducing  a positive relation between GDP growth and foreign debt.
This concern arises as a standard fixed-effects model may capture the effects of miscellaneous valuations of foreign debt, instead of identifying exogenous shocks to foreign debt.
The panel IV analysis using an external instrument will help identify exogenous shocks and reduce the endogeneity concerns.
To this end, we require an instrument that 1) correlates with foreign debt, but 2) is orthogonal to the error term.
	
We choose the CDS spread as an instrument for foreign debt.
The CDS  spread measures the cost for investors to hedge against the sovereign default risk of a country.
Hence, it naturally correlates with foreign debt in the presence of  systemic risks
in financial markets \citep[see][]{ang2013systemic}.
For example, an increase in the CDS spread would lead to a decrease in foreign debt in a country
because investors would reduce demand for securities that are considered more risky.
The CDS spread would be orthogonal to the error term if it is largely determined by global factors. \cite{longstaff2011sovereign} finds that the CDS spread is driven more by global market factors
than by country-specific fundamentals.
\cite{gilchrist2021sovereign} finds that changes in global financial risk account for a substantial portion of the co-movement among the CDS spreads.
Figure \ref{fig:foreign_debt_cds} also shows that the CDS spread co-moves in both country groups especially before 2012. 
Nevertheless, we might still be concerned about a correlation between country-specific fundamentals and the CDS spread.
This is because sensitivity to the global factor can vary across countries depending on country-specific factors.
Considering the possible correlation between the country-specific factors and the CDS spread,
we include macroeconomic fundamentals as control variables.
Assuming that the CDS spread contains information of non-linear combination of global and country-specific factors,
our identification strategy follows the Bartik instrument approach,
which uses the interaction term of global shocks and country-specific exposure as an instrument.\footnote{See \cite{goldsmith2020bartik} and \cite{borusyak2022quasi} for details.}

We employ two-stage least squares (2SLS).
In the first stage we regress foreign debt on the CDS spread and controls to obtain the fitted values of foreign debt.
The CDS spread contains supply-side information such as the willingness of global investors to invest in a country. 
Therefore, as we argued above, the fitted values estimated in the first stage can be interpreted as changes in foreign debt caused by exogenous supply-side shocks.
In the second stage, we regress GDP growth on the fitted values.
Hence, our panel IV regression provides structural interpretation of estimation results identifying the supply-push effects of foreign debt on GDP growth.  

	\begin{figure}[ht!]
	\begin{center}
		\centering
		\includegraphics[width=.55\textwidth]{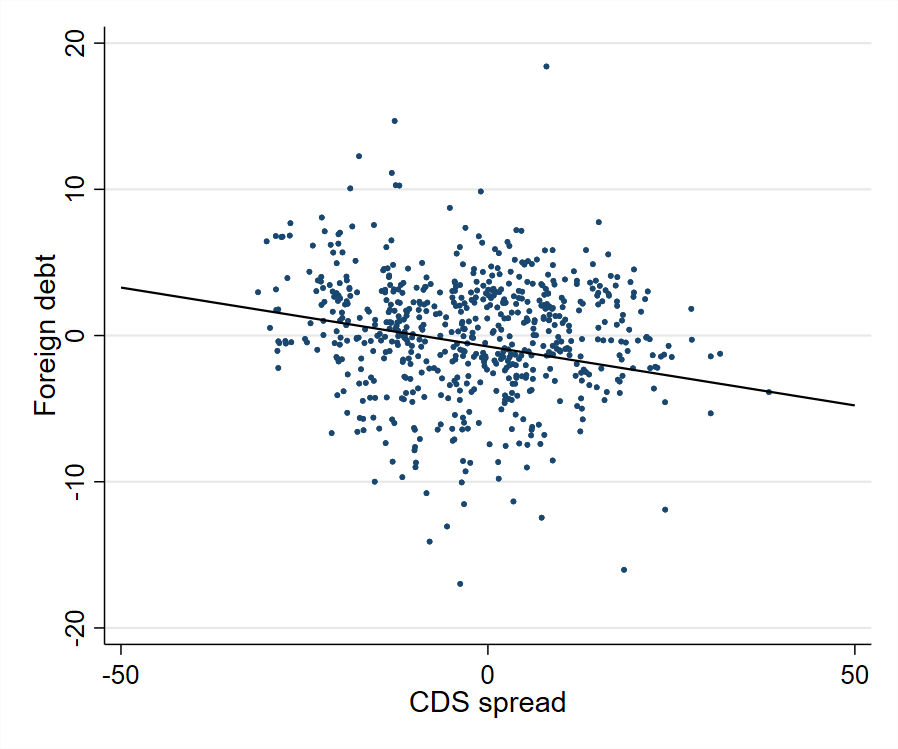}
		\caption{Partial Correlation between CDS Spread and Foreign Debt.}
		\label{fig:partial_first_stage}
	\end{center}	
	\footnotesize Note: Plotted data in each axis are residuals from OLS regressions of the CDS spread and foreign debt on
	controls and country-fixed effects.
	The figure does not show three extreme observations: Greece in 2012 (CDS=508.4 and foreign debt=-54.6);  Argentina in 2014 (CDS=213.6 and foreign debt=-3.0); Argentina in 2015 (CDS=147.0 and foreign debt=-2.4) for visibility.
	The fitted line is based on all the available observations. The omitted observations are located just around the line.
\end{figure}

Figure \ref{fig:partial_first_stage} shows partial correlation between the CDS spread and foreign debt sketching the first stage regression.
Both variables are residualized by the controls, as well as country-fixed effects.
The scatter plots suggest that the CDS spread predicts foreign debt well: Higher sovereign default risks,
which are not explained by observable factors and fixed effects, are associated with smaller foreign debt. 
We confirm this negative relation in a formal analysis reported in Table \ref{fig:Panel-IV_sub-sample}.
	
Our main model is described as follows:

	\begin{equation}
		\Delta GDP_{i,t}=\beta_1{Foreign\_debt}_{i,t}+\Gamma X_{i,t}+\delta_i+\varepsilon_{i,t}
	\end{equation}
where $\Delta GDP_{i,t}$ is the growth rate of real GDP in country $i$ in year $t$;
$Foreign\_debt_{i,t}$ is the debt instrument of portfolio investment in country $i$ in year $t$;
which is instrumented with the CDS spread;
$X_{i,t}$ is a vector of controls; $\delta_i$ captures country-fixed effects; and $\varepsilon_{i,t}$ is the error term.

	\begin{table}[ht!]
	\begin{center}
		\centering
		\includegraphics[width=.7\textwidth]{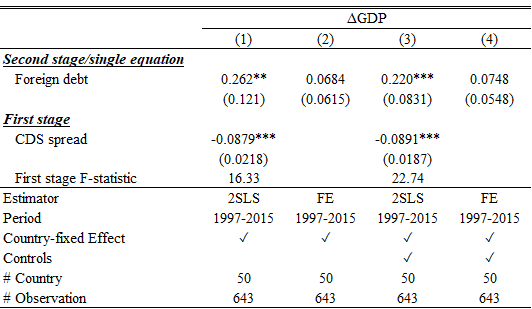}
		\caption{Panel Instrumental Variable Approach (2SLS).}
		\label{fig:Panel-IV_sub-sample}
	\end{center}	
	\footnotesize Note: The dependent variable is real GDP growth. The independent variable is foreign debt (the debt instrument of portfolio investment divided by GDP), instrumented with the sovereign CDS spread in columns 1 and 3. Control variables are the human capital index, the government consumption to GDP ratio (logged), trade openness (logged), the inflation rate and the 1-year lagged level of GDP per capita (logged). Standard errors in parenthesis are clustered on country. *** p\textless0.01, ** p\textless0.05, * p\textless0.1
	\end{table}

Table \ref{fig:Panel-IV_sub-sample} summarizes the results of the panel IV analysis.	Column 1 shows the results without control variables.
In the first stage, the coefficient of the CDS spread on foreign debt is negative and significant suggesting that a lower sovereign default risk is associated with higher foreign debt as expected and shown in Figure \ref{fig:partial_first_stage}.
More importantly, the coefficient of foreign debt on GDP growth is positive and significant in the second stage.  One percentage point increase in foreign debt leads to a 0.262 percentage point increase in real GDP.
The coefficient is over three times larger than the one estimated by the standard fixed-effect estimation with the same sample shown in column 2.
The larger estimate suggests that there might be a downward bias in the fixed-effect estimation caused by endogeneity.
	
The panel IV analysis shows similar results when we include control variables (columns 3-4). In column 3, the coefficient of foreign debt shows a positive and significant sign, and the size of
the coefficient is similar to the one in column 1.
The coefficient is again much larger than that of the fixed-effect estimation shown in column 4, and the result of the first stage is almost the same as the one in column 1.
Moreover, in the panel IV analysis, the first stage F-statistics is sufficiently large, corroborating the relevance of the CDS spread as an instrument.\footnote{The result holds when we use the panel data covering 66 countries including the outlier countries. However, standard errors get larger and the first stage F-statistics become smaller indicating that the countries generate noisy observations and cause inaccurate results. }
	
To summarize, the panel IV analysis in this section reduces endonegeity concerns in our baseline model and indicates that higher foreign debt, in response to a lower sovereign CDS spread, leads to higher GDP growth.

\subsection{Robustness}	
\label{sec:robustness}
	This section presents robustness checks. First, we control for time-specific common factors.
	Column 1 of Table \ref{fig:Robustness_GDP} reports the result of the model including global GDP growth defined by cross-sectional mean of GDP growth in each year.
	As can be seen, the main results hold. The coefficient of foreign debt is positive and significant in the second stage, and that of the CDS
	spread is negative and significant in the first stage.
	Column 2 reports the result of the model including time-fixed effects instead of global GDP growth, which is a popular method to control time-specific common factors.
	The model delivers a similar result to column 1.  
	Nevertheless, we believe that the model including time-fixed effects is over-controlled.
	This is because our identification strategy relies on the global component of the CDS spread being correlated with foreign debt
	into each country.
	Including time-fixed effects will absorb the key information/variation of the global component, which might weaken the explanatory power of the CDS spread on foreign debt. Consistent with the concern the
	F-statistics of the first stage becomes smaller than in column 1.
	Therefore, we include the global GDP growth to control for time-specific common factors in the following analysis.
	
	Second, we include additional controls.
	In the panel IV regression in Section \ref{sec:IV} we might be concerned that the sovereign CDS spread affects GDP growth through a channel other than foreign debt. In that case, we must include the additional variable capturing the channel in our specification to reduce concerns about omitted variable bias.  Financial crises might present such a channel. 
	When a financial crisis occurs, the sovereign default risk increases as GDP is expected to decrease, which in turn decreases foreign debt. 
	In such a case, the CDS spread is systematically related to GDP growth. Therefore, we follow 
	\cite{laeven2013systemic, laeven2020systemic} to include a banking crisis dummy in the following analysis.
	Similarly, the government debt to GDP ratio might present another channel.
	Countries with higher government debt tend to have a higher sovereign CDS spread and lower GDP growth due to a debt overhang problem \citep[e.g.,][]{reinhart2010growth}.
	Thus, we also include the government gross debt to GDP ratio as an additional control.
	Furthermore, we include the regulatory quality index estimated by \cite{kaufmann2011worldwide} and the growth of the private credit to GDP ratio as additional controls, both of which might correlate with foreign debt and GDP growth. 
	Column 3 of Table \ref{fig:Robustness_GDP} summarizes the result of including the additional controls, showing that our main results hold.
	The CDS spread is negatively correlated with foreign debt in the first stage, and its coefficient is positive and significant in the second stage.	
	
	Third, we use an additional instrument.
	Similar to the CDS spread, the sovereign debt rating may be another valid instrument for foreign debt.
	This is because the rating also captures the sovereign default risk of a country and correlates with foreign debt of the country.
	We perform a panel IV analysis using both the CDS spread and the sovereign rating as instruments,
	which enables us to conduct an over-identification test.
	As shown in column 4 of Table \ref{fig:Robustness_GDP}, the main results hold. More importantly, the over-identification test confirms that we can not reject the null hypothesis that
	the excluded instruments are exogenous, although the first stage F-statistics becomes much smaller.
	This result corroborates the validity of using the CDS spread as an instrument.

	Forth, Column 5 of Table \ref{fig:Robustness_GDP} present results when we use 3-year averaged data instead of annual data.
	Specifically, we re-construct our sample into 6 non-overlapping 3-year periods after taking 3-year moving average of each variable.
	The coefficients of foreign debt remain positively significant, although the first stage F-statistic becomes smaller.
	The analysis indicates that our main results capture a solid relation between foreign debt and GDP growth rather than a short-term spurious one. 
	
	Finally, Column 6 of Table \ref{fig:Robustness_GDP} presents the coefficients of foreign debt in OECD and non-OECD countries.
	The model includes interaction terms of foreign debt and OECD/non-OECD dummy.
	The coefficients are positive and significant in both country groups.
	Interestingly, the coefficient is larger in non-OECD countries compared to that in OECD countries. 
	However, the first stage F-statistics become far smaller than 10, which raises the concern that the analysis does not estimate these
	coefficients properly and there exists estimation bias.
	Estimations using OECD or non-OECD sub-samples deliver the similar results.
	Thus, in what follows we assume  homogeneous effects of foreign debt on economic growth across the two country groups.

		\begin{table}[ht!]
		\begin{center}
			\centering
			\includegraphics[width=1.\textwidth]{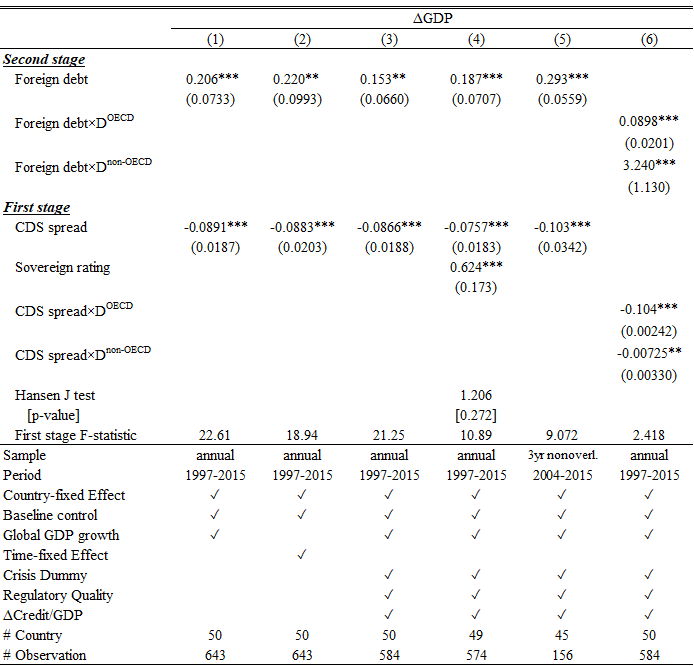}
			\caption{Robustness (2SLS).}
			\label{fig:Robustness_GDP}
		\end{center}	
		\footnotesize Note: The dependent variable is real GDP growth. The endogenous variable is foreign debt (the debt instrument of portfolio investment divided by GDP) instrumented with the sovereign CDS spread (columns 1--4 and 5--6) or the sovereign rating (column 5). In column 6 foreign debt and CDS spread are interacted with OECD or non-OECD dummy. Control variables are the human capital index, the government consumption to GDP ratio (logged), trade openness (logged), the inflation rate, the 1-year lagged level of GDP per capita (logged), the global GDP growth, the banking crisis dummy, the general government gross debt to GDP ratio (logged),  regulatory quality and growth of the private credit to GDP ratio. Standard errors in parenthesis are clustered on country. *** p\textless0.01, ** p\textless0.05, * p\textless0.1
	\end{table}
	
	\clearpage

\section{Contribution of Foreign Debt to GDP Growth in OECD and non-OECD countries\label{sec:mechanism}}

	 The main result presented in Section \ref{sec:growth-debt} is that  an increase in foreign debt, in response to a decrease in the sovereign default risk, leads to an increase in  GDP growth.	
	 We know that the CDS spread is in most years lower in OECD than non-OECD countries (see Figure \ref{fig:foreign_debt_cds}). This suggests that lower sovereign default risks are responsible for higher foreign debt increasing GDP growth
	in OECD countries more than in non-OECD countries. This section verifies the conjecture by calculating  for each county group the contribution of foreign debt, control variables, and residuals by using the coefficients of our model with full set of controls presented in column 3 of Table \ref{fig:Robustness_GDP} and actual values of
	each right-hand-side variables. 
	For example, the contribution of foreign debt is obtained by multiplying actual values of foreign debt in each year
	and each country by the estimated coefficient (=0.153).\footnote{Using different coefficients for OECD and non-OECD countries is another possible option to compare the contribution of
	foreign debt in each group.	Unfortunately, the estimated coefficients are not reliable as discussed in Section \ref{sec:robustness}.}

	\begin{figure}[ht!]
	\begin{center}
		\centering
		\includegraphics[width=.6\textwidth]{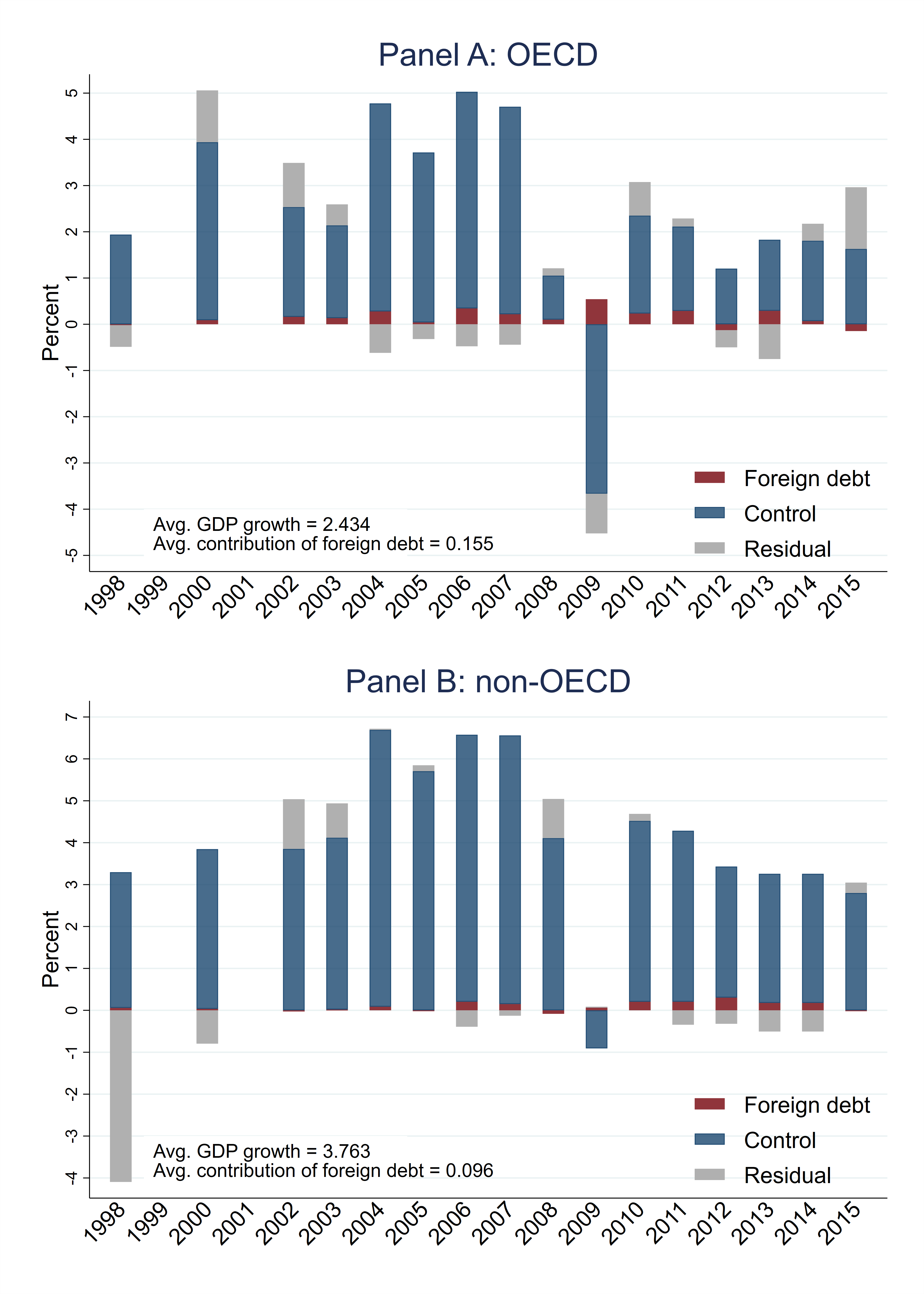}
		\caption{Contribution of foreign debt to GDP growth in OECD and non-OECD countries.}
		\label{fig:growth_accounting}
	\end{center}	
	\footnotesize Note: This figure reports the contribution of foreign debt and control variables based on the coefficients of our model. The values in 1999 and 2001 are missing because of data availability.
\end{figure}

	Figure \ref{fig:growth_accounting} decomposes the GDP growth into three components.
	Panel A shows that the contribution of foreign debt is positive throughout the sample period in OECD countries.
	The average contribution is 0.155 percent, which accounts for 6.4 percent of GDP growth in OECD countries.
	Considering the number of the right-hand-side variables, the estimated contribution is not a small number.
	Interestingly, the contribution of foreign debt becomes the largest in 2009 when the great financial crisis triggered a sharply negative GDP growth.
	This suggests that OECD countries, which are regarded as ``safe havens,''
	benefit from capital inflows even under the crisis.
	In contrast, Panel B shows that the contribution of foreign debt is smaller in non-OECD countries.
	The average contribution is 0.096, which accounts for 2.6 percent of GDP growth in non-OECD countries. Hence, 
	the contribution of foreign debt to economic growth is over two times larger in OECD than non-OECD countries or  foreign debt increases GDP growth in OECD countries (0.155 percentage point) 60 percent more than in non-OECD  countries (0.096 percentage point).

By combing the results in Section \ref{sec:growth-debt} and this section we conclude that lower CDS spread leads to higher foreign debt and contributes more to GDP growth in OECD countries than in non-OECD countries. This holds despite OECD countries having a higher GDP per capita and a slower GDP growth rate than non-OECD countries. 
Therefore, the novel channel of foreign debt and growth we identify via the CDS spread points to the ``real'' effects of the ``exorbitant privilege'' that the US and some other OECD countries have \citep[c.f.,][]{gourinchas2007world}.

\section{The Investment Channel\label{sec:investment}}	

	This section explores the channel through which foreign debt contributes to GDP growth.
	We focus on the investment channel, in which an increase in foreign debt expands the production capacity of a country through accumulation of
	fixed capital.
	Without data that decompose foreign debt into public and private debt, we consider the following possible channels:
	1) private debt owed to foreigners increases private investment,
	2) public debt owed to foreigners increases public investment, and 
	3) public debt owed to foreigners increases private investment.\footnote{We think that it is unlikely  that private debt increases public investment, and therefore exclude the possibility. \cite{alfaro2014sovereigns} provides data that distinguish private and public debt but focus on emerging economies that are a subset of our sample.}
	Channels 1) and 2) are straightforward. Channel 3) considers that public debt might crowd in private investment. For example, private firms may be contracted to carry out a public infrastructure project for which the government raises fund by issuing bonds.  Therefore, there might be a positive relation between public debt and private investment.
	
	To investigate the investment channel, we perform a panel IV analysis with real investment growth as the dependent variable.
	The model is described as follows:

	\begin{equation}
	\Delta Investment^k_{i,t}=\beta_1{Foreign\_debt}_{i,t}+\Gamma X_{i,t}+\delta_i+\varepsilon_{i,t}
	\end{equation}

	where $\Delta Investment^k_{i,t}$ is the growth of gross fixed capital formation in country $i$ in year $t$ with 
	superscript $k$ representing total, private, or public investment, and 	$Foreign\_debt_{i,t}$ indicates the debt instrument of portfolio investment as before. 
	Total investment is measured in a constant local currency unit and provided by the WDI database.
	Private and public investments are provided by the IMF Investment and Capital Stock Database and measured in a constant US dollar unit instead of a local currency unit.
	Following the panel IV analysis presented in Section \ref{sec:growth-debt}, foreign debt is instrumented with the CDS spread and the sample period is from 1997 to 2015.

	\begin{table}[ht!]
	\begin{center}
		\centering
		\includegraphics[width=1\textwidth]{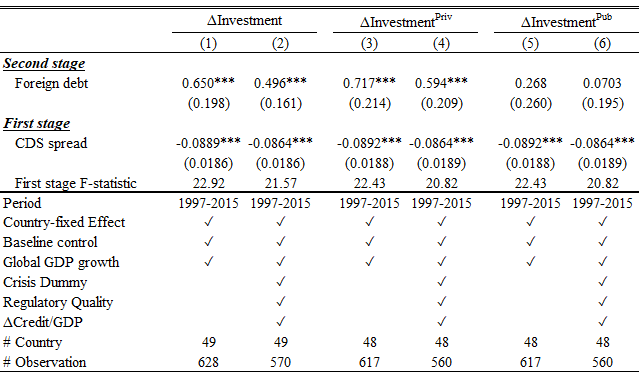}
		\caption{Investment Channel (2SLS).}
		\label{fig:Panel-IV_investment}
	\end{center}	
	\footnotesize Note: The dependent variable is real GDP growth, real private investment growth, or real public investment growth. The endogenous variable is foreign debt (the debt instrument of portfolio investment divided by GDP) instrumented with the sovereign CDS spread. Control variables are the human capital index, the government consumption to GDP ratio (logged),  trade openness (logged), the inflation rate, the 1-year lagged level of GDP per capita (logged), the global GDP growth, the banking crisis dummy, the general government gross debt to GDP ratio (logged), regulatory quality and  growth of the private credit to GDP ratio. Standard errors in parenthesis are clustered on country. *** p\textless0.01, ** p\textless0.05, * p\textless0.1
	\end{table}

	Table \ref{fig:Panel-IV_investment} summarizes the results. Column 1 shows that the coefficient of foreign debt on total investment is positive and significant. One percentage point increase in foreign debt leads to a 0.650 percentage point increase in total investment growth. The coefficient is much larger than that of GDP growth shown in columns 1 and 3 of Table \ref{fig:Panel-IV_sub-sample}.
	The result remains unchanged when we include additional controls as shown in column 2.
	Hence, the results show that an increase in investment, in response to an increase in foreign debt, leads to an increase in  GDP growth.
	Columns 3--6 of Table \ref{fig:Panel-IV_investment} decompose the total investment into private and public investments. The coefficients of foreign debt on private investment are positive and significant (columns 3 and 4). This result is consistent with that of total investment growth in columns 1 and 2.
	Moreover, it also supports the channel of a possible crowding-in of private investment when public debt increases.
	In contrast, the coefficients on public investment are much smaller and insignificant (column 5 and 6) possibly due to the counter-cyclical nature of fiscal stimulus packages.

	To summarize, we show that foreign debt increases private investment. If foreign debt is largely private, the link to private investment is straightforward. If it is largely public, government procurement of goods and services may still increase private investment.%
	\footnote{It is well-known that companies such as Lockheed Martin for defense, Boeing for aircraft, Amazon, IMB and Microsoft for information technology, to just name a few, are large contractors of the US government.} Moreover, even without this direct link, foreign debt might increase private investment if foreign debt correlates positively with private debt.

\section{The Dynamic Effects of Foreign Debt\label{sec:dynamic-effects}}

	The analysis presented in the previous sections focuses on the contemporaneous effects of foreign debt
	on both GDP and investment growth. To investigate the dynamic effects on the
	subsequent growth of both GDP and investment, the analysis in this section combines the local projection approach developed by \cite{jorda2005estimation} with IV methods (i.e., LP-IV approach).
	The model is described as follows:
	\begin{multline}
		\Delta_{h} GDP_{i,t+h}=\ln{GDP_{i,t+h}}-\ln{GDP_{i,t-1}}=\beta^h{Foreign\_debt}_{i,t}+\Gamma^hX_{i,t}+\Phi^hW_{i,t-1}\\[1.ex]
		+\delta^h_i+\varepsilon^h_{i,t+h}
	\end{multline}
	where $\Delta_{h} GDP_{i,t+h}$ is the cumulative growth of real GDP in country $i$ from year $t-1$ to $t+h$ (for $h=0, 1, 2$, and 3) and $Foreign\_debt_{i,t}$ indicates the debt instrument of portfolio investment as before, which is instrumented with the CDS spread.
	Following \cite{jorda2015betting} and \cite{stock2018identification}, the model includes current and lagged control variables ($X_{i,t}$ and $W_{i,t-1}$) used in Table \ref{fig:Robustness_GDP} and \ref{fig:Panel-IV_investment} as well as the lagged dependent variable, the lagged endogenous variable, and the lagged instrument (i.e., lagged GDP growth, foreign debt, and the CDS spread).

	The sequence of coefficients $\beta^h$ captures the impulse response of GDP
	to an increase in foreign debt. In the local projections, we fix foreign debt on the right-hand-side in year $t$, and estimate real GDP growth on the left-hand-side into the future.
	For example, with $h=0$, $\beta^{0}$ is the effect of an increase in foreign debt in year $t$ on GDP growth from year $t-1$ to $t$. Thus, it captures the contemporaneous effect of foreign debt on GDP growth.
	Similarly, with $h=2$, $\beta^2$ will capture the effect of an increase in foreign debt in year $t$
	on the cumulative GDP growth from year $t-1$ to $t+2$.

	\begin{figure}[ht!]
	\begin{center}
		\centering
		\includegraphics[width=1\textwidth]{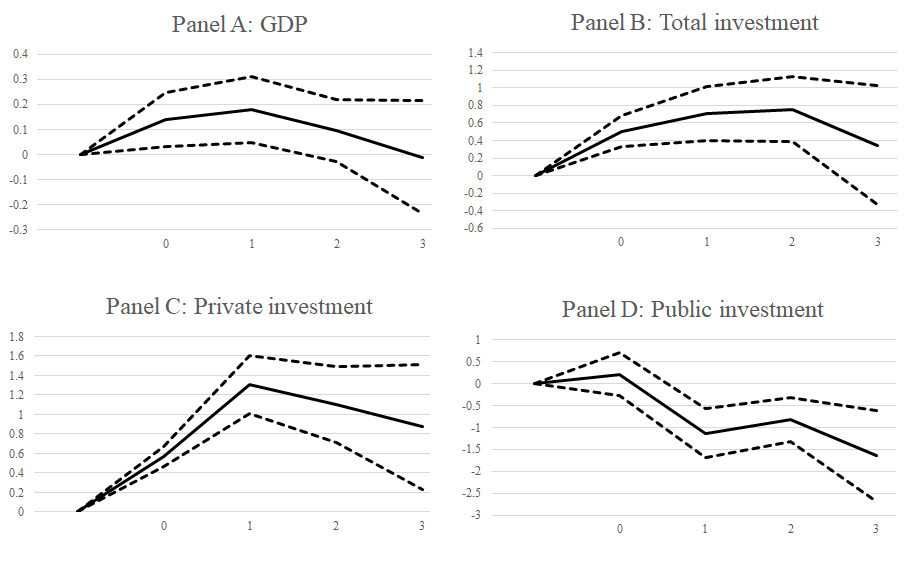}
		\caption{Impulse Response of Real GDP and Investment Growth (LP-IV).}
		\label{fig:LP-IV_IRFs}
	\end{center}	
	\footnotesize Note: The solid line represents the response of the dependent variable to an increase in foreign debt for forecast horizon h=0, 1, 2, and 3. Dashed line represents the 95\% confidence interval calculated based on the standard error clustered on country. The horizontal axis represents the year after an increase in foreign debt. The dependent variable is cumulative growth of real GDP (panel A), total investment (panel B), private investment (panel C), or public investment (panel D) Control variables are the current and lagged values of the human capital index, the government consumption to GDP ratio (logged), trade openness (logged), the inflation rate,  the level of GDP per capita (logged), global GDP growth, the banking crisis dummy, the general government gross debt to GDP ratio (logged), regulatory quality and growth of the private credit to GDP ratio. Specification also includes the lagged dependent variable, foreign debt, the sovereign CDS spread and country-fixed effects.
	\end{figure}

	Panel A of Figure \ref{fig:LP-IV_IRFs} shows the impulse response of GDP to an increase in foreign debt.
	The dynamic effect of foreign debt is positive and significant over years, but it peaks out after one year from the shock.
	The result indicates that an increase in foreign debt leads to an increase in the level of GDP and the positive effect remains three years after the shock.
	Panel B of Figure \ref{fig:LP-IV_IRFs} shows the impulse responses of total investment
	to an increase in foreign debt.
	The dynamic effect is positive and significant and then becomes smaller and insignificant three years after the shock supporting the investment channel.
	Panels C and D of Figure \ref{fig:LP-IV_IRFs} show the dynamic effect
	when total investment is decomposed into private and public investments.
	Panel C shows that the dynamic effect on private investment is largely similar to that in panel B of Figure \ref{fig:LP-IV_IRFs}.
	The dynamic effect of foreign debt is first positive and significant but peaks out after two years from the shock,
	which is consistent with the dynamics of cumulative GDP growth (Panel A).
	In contrast, the dynamic effect on public investment is much smaller or negative (panel D of Figure \ref{fig:LP-IV_IRFs}).
	Therefore, we conclude that an increase in 
	foreign debt leads to an increase in private investment and GDP growth and the dynamic effect through capital accumulation lasts for two or three years.

\section{Conclusion\label{sec:conclusion}}

	We find a positive relationship between foreign debt and GDP growth. Using the sovereign default risk as an instrumental variable we find that the relationship is causal in nature. Moreover, using a local projection analysis we find that an increase in foreign debt leads to an increase in investment and then GDP growth in subsequent years. On average, the sovereign default risk is lower and foreign debt is higher in OECD than non-OECD countries. This suggests that higher foreign debt contributes to GDP growth more in OECD than non-OECD countries. We verify this claim by estimating the contribution of foreign debt to GDP growth based on our model.
	Given that OECD countries grow slower on average than non-OECD countries but issue most of foreign debt, our findings indicate that the low sovereign default risk gives the US and other OECD countries the exorbitant privilege that contributes to their GDP growth.

\clearpage

\appendix
\counterwithin{figure}{section}
\counterwithin{table}{section}

\section{Appendix}
\label{sec:appendix}

	\begin{table}[ht!]
	\begin{center}
		\centering
		\includegraphics[width=.8\textwidth]{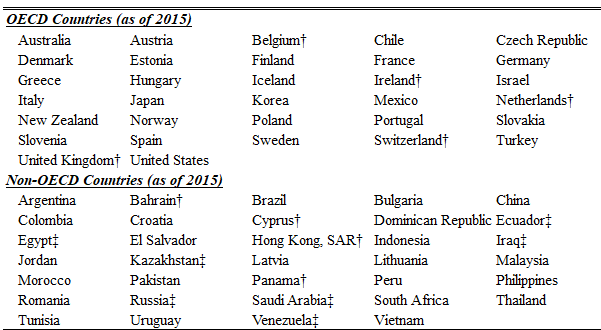}
		\caption{List of countries.}
		\label{fig:country_list}
	\end{center}	
	\footnotesize Note: $\dagger$Financial center (\cite{lane2018external}). \ddag Oil exporter, oil rents (\% of GDP, average of 1980-2015) exceed 8\%.
	\end{table}

	\begin{table}[ht!]
	\begin{center}
		\includegraphics[width=0.6\textwidth]{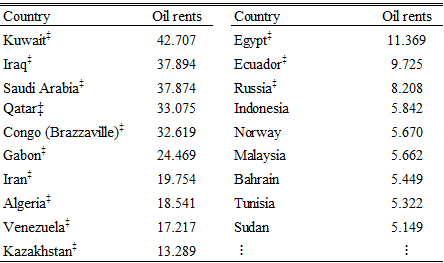}
		\caption{Ranking of Oil Rents (\% of GDP)}
		\label{fig:Ranking_oil}
	\end{center}	
		\footnotesize  Note: This table reports the ranking of average oil rents (\% of GDP) from 1980 to 2015. \ddag Oil exporter. Oil rents are the difference between the value of crude oil production at world prices and total costs of production
	\end{table}

\clearpage

\bibliographystyle{ecta}

\bibliography{international_debt_flows_bib}

\end{document}